\documentstyle[aps,prl]{revtex}

\begin{document}
\draft
\title{Tests of Bose-broken Symmetry in\\
Atomic Bose-Einstein Condensates.}
\author{T. Wong, M. J. Collett, S. M. Tan, and D. F. Walls}
\address{Department of Physics, University of Auckland, \\
Private Bag 92019, Auckland, New Zealand.}
\author{E. M. Wright}
\address{Optical Sciences Center\\
University of Arizona\\
Tucson, AZ 85721, USA}
\maketitle

\begin{abstract}
We present an elementary model of the collapses and 
revivals in the
visibility of the interference between two atomic 
Bose-Einstein condensates.
We obtain different predictions of the revival times 
whether we conserve or
break atom number conservation from the outset. The 
validity of Bose-broken
symmetry can be tested by observations of these collapses 
and revivals.
\end{abstract}

\pacs{PACS numbers: 03.75 Fi, 05.30 -d}

The recent experimental realization of a weakly interacting 
Bose-Einstein
condensate in an alkalic gas \cite{anderson,bradley,davis} 
has stimulated
considerable theoretical work on the properties of these 
condensates. One
question which has received much attention concerns the 
phase of the
condensate and how it is established \cite
{Jan,Nar,Cas,Cir,MeasureScheme,Wong,Bur,Molmer}. A 
conventional approach is
to invoke Bose-broken symmetry arguments \cite{Gri93} and 
select an
arbitrary phase from an initial state which has a random 
phase. (It is not
possible to measure the absolute phase of a condensate so 
that the phases we
talk about are the relative phases between two 
condensates.) A condensate
formed in the ground state of a trap will be in a number 
state, though we
may not have knowledge of what this number is. A number 
state may be
considered as a continuous superposition of coherent states 
all with
different phases; spontaneous symmetry breaking then 
selects just one
(arbitrary) phase. Recently it has been shown that a 
relative phase may be
established between two condensates initially in number 
states by
measurements \cite{Jan,Nar,Cas,Cir,MeasureScheme,Wong}. 
This leads to the
conclusion that considering the condensate to be in either 
a coherent state
or a number state are equivalent. Note that whereas atom 
number is conserved
in the number state case, an initial coherent state is a 
Poissonian
superposition of number states and hence the atom number 
conservation is
broken.

In this paper we consider the evolution of the visibility 
in the
interference between two Bose-Einstein condensates in the 
presence of
collisions. Collisions give rise to the collapse and 
revival of the
macroscopic wave-function of small atomic condensates 
composed of $%
10^{3}-10^{5}$ atoms \cite{Wright}, which results in 
collapse and revival in
the visibility when two condensates are interfered. The 
central result of
this paper is that the revival time is strongly dependent 
(a factor of 2) on
the initial state chosen for the condensate. This gives us 
a way to
determine unambiguously the true quantum state of the 
condensate. We shall
illustrate our results for three different initial states 
of the two
condensates. In the first example we invoke Bose-broken 
symmetry and
consider the product of two coherent states so that number 
conservation is
broken. In the other two examples we consider an entangled 
state of the two
condensates with fixed total number $N$ so that no 
assumption of Bose-broken
symmetry is invoked. The evolution of the condensates under 
the influence of
collisions is studied for each of the above initial 
conditions and the
collapse and revival times are compared.

Our model comprises two Bose-Einstein condensates which 
both occupy the
ground-state of their respective traps and are described by 
the atom
annihilation (creation) operators $\hat{a}$ ($\hat{a}
^{\dagger }$) and $\hat{%
b}$ ($\hat{b}^{\dagger })$. Atoms are released from each 
trap with momenta $%
{\bf k}_{1}$and ${\bf k}_{2}$ respectively, producing an 
interference
pattern which enables a relative phase to be measured.

The intensity of the atomic field is given by 
\begin{equation}
I\left( {\bf x},t\right) =I_{0}\left\langle \left[ \hat{a}
^{\dag }\left(
t\right) e^{i{\bf k}_{1}\cdot {\bf x}}+\hat{b}^{\dag }
\left( t\right) e^{i%
{\bf k}_{2}\cdot {\bf x}}\right] \left[ \hat{a}\left( 
t\right) e^{-i{\bf k}%
_{1}\cdot {\bf x}}+\hat{b}\left( t\right) e^{-i{\bf k}_{2}
\cdot {\bf x}%
}\right] \right\rangle ,
\end{equation}
where $I_{0}$ is the single atom intensity. Atoms within 
each condensate
collide and this may be described by the Hamiltonian 
\begin{equation}
H=\frac{1}{2}\hbar \chi \left[ \left( \hat{a}^{\dag }\hat{a}
\right)
^{2}+\left( \hat{b}^{\dag }\hat{b}\right) ^{2}\right] ,  
\label{collisions}
\end{equation}
where $\chi $ is the collision rate between the atoms 
within each
condensate. Cross-collisions between the two condensates, 
described by the
term $\hat{a}^{\dag }\hat{a}\hat{b}^{\dag }\hat{b},$ are 
not included since
they are dependent on the actual geometry of the physical 
situation. The
coefficient of this term could be anywhere between zero and 
$\hbar \chi $
depending on the overlap between the two condensates. We 
consider the case
where this overlap is small.

Including the time dependence of $\hat{a}$ and $\hat{b}$ 
due to the
collisions described by Eq. (\ref{collisions}), we get for 
the intensity 
\begin{equation}
I\left( {\bf x,}t\right) =I_{0}\left\{ \left\langle \hat{a}
^{\dag }\hat{a}%
\right\rangle +\left\langle \hat{b}^{\dag }\hat{b}
\right\rangle
+\left\langle \hat{a}^{\dag }\exp \left[ i\left( \hat{a}
^{\dag }\hat{a}-\hat{%
b}^{\dag }\hat{b}\right) \chi t\right] \hat{b}\right\rangle 
e^{-i\phi \left( 
{\bf x}\right) }+h.c.\right\} ,  \label{intensity}
\end{equation}
where $\phi \left( {\bf x}\right) =\left( {\bf k}_{2}-{\bf 
k}_{1}\right)
\cdot {\bf x.}$ The third term in the expression for 
$I\left( {\bf x,}%
t\right) $ gives rise to interference fringes. However the 
interference
pattern will be modified in time due to the decohering 
effects of the
collisions.

We shall now proceed to demonstrate how the interference 
pattern evolves for
different initial states of the condensates. We will 
consider two classes of
initial states, one which uses Bose-broken symmetry 
arguments and the other
which does not. This Bose symmetry is associated with atom 
number
conservation, so breaking this symmetry allows us to write 
the wave-function
as a superposition of number states. This symmetry is 
deemed to be broken in
the formation of the condensate. The uncertainty in the 
number leads to the
wave-function possessing a definite phase via the 
number-phase uncertainty
relation $\Delta n\Delta \phi \sim 1.$ Following this line 
of argument one
can imagine an initial fixed number of atoms before the 
condensate is
formed. This has indeterminate phase since it can be 
expressed as a
continuous superposition of coherent states each possessing 
a particular
phase. When the condensate is formed, the symmetry is 
broken with one of the
coherent state selected, its phase becomes the phase of the 
condensate. In
the spirit of Bose-broken symmetry we then take each 
condensate to be
initially in a coherent state 
\begin{equation}
\left| \varphi _{B}\right\rangle =\left| \alpha 
\right\rangle \left| \beta
\right\rangle .
\end{equation}
This yields for the intensity in Eq. (\ref{intensity}) 
\begin{equation}
I\left( {\bf x},t\right) =I_{0}\left| \alpha \right| ^{2}
\left\{ 1+\exp
\left[ 2\left| \alpha \right| ^{2}\left( \cos \chi 
t-1\right) \right] \cos
\left[ \phi \left( {\bf x}\right) -\phi \right] \right\}  
\label{Int_coh}
\end{equation}
where we have set the amplitude of the two condensates to 
be equal to
maximize the visibility of the interference pattern. The 
relative phase
between the two condensates is defined to be $\phi ,$ so 
that the
relationship between the complex amplitudes is $\beta 
=\alpha e^{-i\phi }.$
The exponential term describes the time dependence of the 
visibility of the
interference pattern. Inside this exponential, we have a 
periodic function
of period $2\pi /\chi ,$ corresponding to revival times 
where the visibility
is $1$. This visibility suffers a minimum half-way between 
these revivals
with a value of $\exp \left( -4\left| \alpha \right| ^{2}
\right) $; it
varies smoothly in time from these minima to their local 
maxima.

In our second class of initial states we will not use 
Bose-broken symmetry,
so we conserve the total atom number $N$ of the two 
condensates. We shall
consider two entangled states of the condensates with fixed 
total number $N.$
In the first example we consider the product of two 
coherent states $|\alpha
\rangle \otimes |\beta \rangle $ projected onto a number 
state basis\cite
{Molmer}. This basis is truncated to size $N$ with equal 
amplitudes $\left|
\alpha \right| =\left| \beta \right| =\sqrt{N/2}.$ We can 
define a relative
phase between the condensates by superposing number 
difference states. This
entangled state is 
\begin{equation}
|\varphi _{N}\rangle =2^{-N/2}e^{iN\phi }\sqrt{N!}\sum_{k=0}
^{N}\frac{%
e^{-ik\phi }}{\sqrt{k!\left( N-k\right) !}}|k\rangle \,
\otimes |N-k\rangle
,\,  \label{molmer}
\end{equation}
where $\phi $ is the relative phase between the 
condensates. Note how each
entangled number state has fixed total atom number $N$ and 
number difference 
$2k.$ The intensity for this initial state is 
\begin{equation}
I\left( {\bf x},t\right) =I_{0}\frac{N}{2}\left\{ 1+\cos 
^{N-1}\chi t\cos
\left[ \phi \left( {\bf x}\right) -\phi \right] \right\} .  
\label{Int_ent}
\end{equation}
The visibility of the interference pattern is $\left| \cos 
^{N-1}\chi
t\right| $. The parity of the total atom number $N,$ 
whether it is odd or
even, plays a role in the revivals. When $N$ is odd, the 
cosine term is
raised to an even power giving a revival period of $\pi 
/\chi $ since it is
never negative. When $N$ is even, the cosine term is raised 
to a odd power
also giving a revival period of $\pi /\chi $ but with each 
alternate revival
occurring with the phase shifted by $\pi $ radians. In both 
cases, the
revival period is one half of the period predicted in the 
previous case
where we used Bose-broken symmetry whatever the parity of 
$N$ is.

The significant difference between the two cases is that 
the total number $N$
is not fixed for the case of Bose-broken symmetry whereas 
it is for the
other case. The factor of two difference in the period can 
be explained by
looking at the exponential term in Eq. (\ref{intensity}). 
Inside this
exponential we have the atom number difference operator 
$\hat{a}^{\dag }\hat{%
a}-\hat{b}^{\dag }\hat{b}$ which is quantized in units of 
$2$ when the total
number is fixed and units of $1$ when it isn't fixed. Thus 
we have either an 
$\exp \left( i2n\chi t\right) $ term or an $\exp \left( 
in\chi t\right) $
term, where $n$ is an integer. This gives rise to the 
factor of two
difference in the period.

As a third example we shall consider an initial state which 
may be formed by
quantum measurement. The state formed from two condensates 
initially in
number states $N/2$ after $m$ atoms have been detected is 
of the general form

\begin{equation}
|\varphi _{m}\rangle =\sum_{k=0}^{m}c_{k}|n-m+k,n-k\rangle ,
  \label{wfn0}
\end{equation}
which is normalized so that the coefficients satisfy the 
relation $\sum
\left| c_{k}\right| ^{2}=1,$ with the values of the $c_{k}$ 
depending on the
actual sequence of measurements. Note that this prepared 
state has a fixed
total number of atoms ($2n-m$ atoms). In order to study the 
collapse and
revival of this coherence we need to include the decohering 
effects of
collisions. A Monte-Carlo wave-function method has been 
used to simulate
numerically the time evolution between subsequent atom 
position measurements
including the effects of collisions 
\cite{MeasureScheme,Wong}. The model we
have used so far already include collisions so all that we 
need to do is to
consider the prepared state $|\varphi _{m}\rangle $ as our 
third description
of the wave-function for the two condensates. Using the 
expression for the
intensity given by Eq. (\ref{intensity}), we obtain 
\begin{equation}
I\left( {\bf x,}t\right) =I_{0}\left\{ 2n-m+\sum_{k=1}^{m}
c_{k}^{*}c_{k-1}%
\sqrt{\left( n-k\right) \left( n-m+k-1\right) }\exp \left[ 
i\left(
2k-m-1\right) \chi t-i\phi \left( {\bf x}\right) \right] 
+h.c.\right\} .
\end{equation}
Any collisional effects occurring during the state 
preparation, which we
have not included here, will affect the coefficients of 
$|\varphi
_{m}\rangle $. For a good state preparation in the sense 
that the entangled
state possesses high coherence we will expect that the 
relative phases
between the neighboring coefficients $c_{k}$ are very 
similar and in fact
give a good estimate of the relative phase between the two 
condensates.
Looking back at the coefficients previously defined in the 
second
description of the wave-function, Eq. (\ref{molmer}), we 
see that the phase
between the neighboring coefficients is identical for 
$|\varphi _{N}\rangle
. $ By comparing the coefficients of the prepared state 
$|\varphi
_{m}\rangle $ in numerical simulations where small 
collisional rates have
been included (see reference \cite{Wong}), we know that 
there is a fairly
well defined phase between neighboring entangled number 
states in $|\varphi
_{m}\rangle $ so that we may write, to a good approximation,
 $%
c_{k}^{*}c_{k-1}=A_{k}e^{i\phi },$ where $\phi $ is the 
relative phase
between the condensates. The interference pattern then 
becomes 
\begin{equation}
I\left( {\bf x,}t\right) =I_{0}\left\{ 2n-m+\sum_{k=1}^{m}
{\cal A}_{k}\cos
\left[ \left( 2k-m-1\right) \chi t+\phi -\phi \left( {\bf x}
\right) \right]
\right\} ,
\end{equation}
where 
\begin{equation}
{\cal A}_{k}=A_{k}\sqrt{\left( n-k\right) \left( 
n-m+k-1\right) },
\end{equation}
which consists of a summation over cosines weighted by 
${\cal A}_{k}$ with
differing multiple frequencies of $\chi t$. Separating the 
time dependence
from the phase $\phi $ and the position $\phi \left( {\bf x}
\right) $ terms
in the cosine, we obtain 
\begin{equation}
I\left( {\bf x,}t\right) =I_{0}\left[ 2n-m+\sum_{k=1}^{m}
{\cal A}_{k}\cos
\left[ \left( 2k-m-1\right) \chi t\right] \cos \left[ \phi 
\left( {\bf x}%
\right) -\phi \right] \right] .
\end{equation}
$2k-m-1$ is odd (even) whenever $m$ is even (odd). Since 
the initial total
number is $2n$ which by definition is even, the parity of 
$m$ determines the
parity of the total atom number of the entangled state. 
When the parity is
even, the intensity consists of a constant term plus a 
weighted sum over
cosines with frequencies which are in multiples of $2\chi 
t$ and thus the
period is $\pi /\chi .$ On the other hand, if the parity is 
odd we have a
weighing over cosines with frequencies which are odd 
multiples of $\chi t.$
The ``visibility'' of the interference pattern $I\left( 
t\right) $ would
then be $-1$ or $1$ for odd or even multiples of the 
revival time $\pi /\chi 
$ respectively. Visibility in the usual sense is always 
positive; the change
in sign indicates a $\pi $ phase shift in the interference 
pattern. Thus
these two examples of entangled states which preserve atom 
number agree on
the period and parity properties of the collapses and 
revivals.

The visibilities of the interference patterns when 
Bose-broken symmetry is
invoked, Eq. (\ref{Int_coh}) and when it is not, Eq. 
(\ref{Int_ent}) are
shown in Fig. (\ref{fig01}a) and (\ref{fig01}b). We have 
not shown a graph
of the third description because it is implicitly a 
measurement scheme which
establishes phase and would give a result very similar to 
Eq. (\ref{fig01}%
b). In order to verify experimentally the dependence of the 
visibility on
time shown in Fig. (\ref{fig01}), the following procedure 
may be employed.
Initially a sequence of measurements of the output field is 
made in order to
establish a relative phase between the condensates. The 
position of the
peaks of this initial interference pattern is recorded. The 
condensates are
then left to evolve freely under the influence of the 
Hamiltonian Eq. (\ref
{collisions}). After a time $t$ of free evolution, a second 
sequence of
measurements of the output field is made. This establishes 
a new
interference pattern with peaks located at positions which 
may differ from
those of the initial pattern. The calculated visibility 
refers to the
distribution of the position of the peaks of the new 
interference pattern
relative to those of the initial pattern. As such it can 
only be determined
by several pairs of measurement sequences which 
alternatively establish and
probe the phase, separated by the time interval $t.$

The validity of imposing Bose-broken symmetry from the 
outset as a means of
describing the quantum dynamics of Bose-condensed systems 
can be tested by
measuring the visibility of the interference pattern 
between two condensates
as a function of time. For small atomic samples composed of 
$10^{3}-10^{5}$
atoms the visibility undergoes a series of collapses and 
revivals. The
magnitude of the revival period of the visibility will test 
whether the
symmetry is in fact broken in the condensation process or 
whether a quantum
state involving a fixed number of atoms is more 
appropriate. Indirect
measurements are also possible from recent light scattering 
proposals
suggested by Imamo\={g}lu and Kennedy \cite{ImaKen96} and 
Javanainen \cite
{Jav96} using independent condensates coupled to a common 
excited state. The
recent production of two overlapping condensates of $^{87}
$Rb in different
spin states has increased the experimental feasibility of 
these proposals..

This research was supported by the University of Auckland 
Research
Committee, the New Zealand Lottery Grants Board and the 
Marsden Fund of the
Royal Society of New Zealand. We would like to thank Prof. 
R. Graham for
interesting comments.

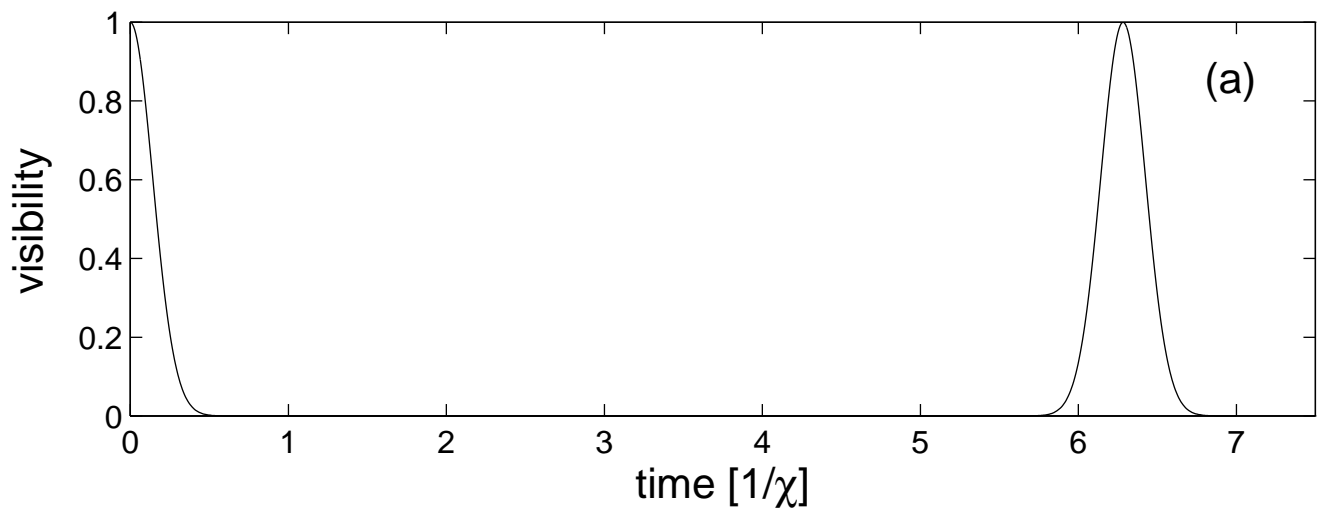
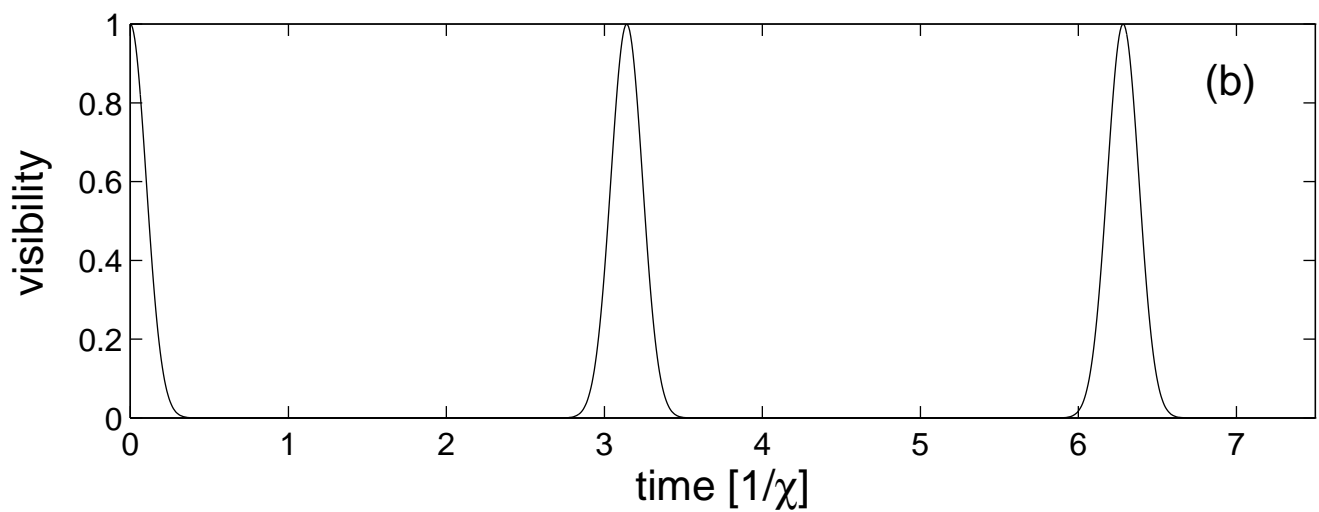
\end{document}